# Treatise on the Resolution of the Diamond Problem After 200 Years

Reginald B. Little[*] and Joseph Roache

Florida A&M University

Florida State University

National High Magnetic Field Laboratory

Tallahassee, Florida 32308

**Abstract:**
The problem of the physicochemical synthesis of diamond spans more than 200 years, involving many giants of science. Many technologies have been discovered, realized and used to resolve this diamond problem. Here the origin, definition and cause of the diamond problem are presented. The Resolution of the diamond problem is then discussed on the basis of the Little Effect, involving novel roton-phonon driven (antisymmetrical) multi-spin induced orbital orientation, subshell rehybridization and valence shell rotation of radical complexes in quantum fluids under magnetization across thermal, pressure, compositional, and spinor gradients in both space and time. Some experimental evidence of this magnetic quantum Resolution is briefly reviewed and integrated with this recent fruitful discovery. Furthermore, the implications of the Little Effect in comparison to the Woodward-Hoffman Rule are considered. The distinction of the Little Effect from the prior radical pair effect is clarified. The better compatibility of radicals, dangling bonds and magnetism with the diamond lattice relative to the graphitic lattice is discussed. Finally, these novel physicochemical phenomena for the Little Effect are compared with the natural diamond genesis.

* corresponding author; email : redge_little@yahoo.com

## 1. Introduction – The Diamond Problem:

A long history of giants of science has defined and contributed to the solution of the diamond problem. Section 2 considers the cause of the diamond problem. The list includes Newton, Boyle, Lavoisier, Guillton, Clouet, Karazin, Despretz, Hannay, Moisson, Roozeboom, Jessup, Rossini, Tammann, Parson, Einstein, Leipunski, Bridgman, Berman, Simon, Hershey, von Platen, Lundblad, Hall, Bundy, Strong, Wentorf, Cannon, Eversole, Deryagin, Angus, Fedoseev, Setaka, Matsumoto, Yugo, Linarres, Hemley, Sumiya, and Little. On the basis of these investigators, many technologies have been considered for resolving the diamond problem. These technologies include electric arcs; metal solvents; metal catalysts; electric ovens; electric resistive heaters; anvil vices and presses; electron beams; atomic beams; x-rays, alpha, and beta irradiators; exploding media; rapid lasing and liquid nitrogen quencher; chemical vapor deposition (CVD); hydrogenous plasma and microwave apparatuses; hot filaments; flames; and **recently strong magnets**. Some of these technologies have resulted in partial successes by the direct method, the indirect method and the H plasma metastable method of forming diamond. The complete Resolution here considers all three of these methods along with natural diamond formation and demonstrates complete Resolution by external magnetization (using external magnetization in conjunction with these three methods and considering intrinsic magnetism in mantle Kimberlite).

On this basis, the synthesis and understanding of diamond have developed along side developments in chemistry, physics and engineering during the last 300 years. In particular, the development of quantum theory in the last century and its detailed applications to matter are realized and put forth in this Resolution for a complete solution to the diamond problem, thereby providing a giant leap to solve the last piece of the puzzle of the diamond problem. The puzzle is Resolved here by determining and manipulating the nonclassical and high spin importance of the intermediary phases (during the direct; the indirect; and H plasma processes) associated with the pertinent activated reaction trajectories for diamond formation by realizing and controlling the high spin, magnetic, quantum, and fluidic natures of these intermediary phases during these diamond forming processes. Section 3 considers this Resolution in details. This nonclassical nature of the diamond forming reaction trajectories follows from the low density of electronic states of carbon atoms and lack of inner core p subshell such that multi-atom quanta interactions are required for diamond symmetry and such large quanta and specific momenta of multi-atoms persist beyond atomic and molecular into mesoscopic and into macro-dimensions. However, such low densities of states are characteristic of Bose-Einstein statistics for the quantum fluidic intermediates. Here higher densitites of electronic states are realized by fermionic quanta fluidic intermediates for different Fermi-Dirac statistical dynamics. Therefore, the nonclassical reaction trajectories and Bose-Einstein statistics result in the large quanta and large activation energies. But the Resolution put forth here by magnetization and many body Interactions and consequent Fermi Dirac statistics result in smaller quanta and lower activation barriers. Such fermionic, magnetic, quantum, fluidic conditions and media provide chemical energy, chemical volume (permittivity-permeability) and chemical pressures to lower thermo-mechanical volume, and pressure requirements of activation for forming diamond. The Resolution determines the formation of these magnetic, quantum fluids for all processes that form diamond: the direct (liquid carbon), the indirect (liquid ferrometals), and the metastable (atomic H-plasma) processes. The synthesis conditions of higher pressures, higher temperatures and thermal plasma induce intrinsic magnetic properties in the growth media for fermionic statistics. These intermediary magnetic fluids behave nonclassically on microscopic and collective scales and in different ways relative to the more classical fluids like air and water.

Here the Resolution of the Diamond Problem is determined based on further taking advantage (via external magnetization) of these intrinsic magnetic nonclassical aspects of the carbonaceous (C), hydrogenous (H), and ferrometal (M) quantum fluids that define and lower the reaction pathways that form diamond. Intrinsic magnetic aspects of such C, M, and H quantum fluids are identified as important to nucleate and grow diamond. In particular, magnetic characteristics of the C, M, and H fluidic solvents and magnetic solute complexes ($CC_4 \bullet C_x$), ($CC_4 \bullet M_x$) and ($CC_4 \bullet H_x$) are determined as important states along the reaction trajectories causing fruitful and unique dynamics of carbonaceous intermediates for $sp^3$ carbon formation, stabilization, orientation, organization, correlation, and knitting into diamond. The C, M and H quanta fluidic solvent and the various carbonaceous complex solutes are reactants and products of many radical chemical reactions for novel chemical dynamics. These diamond forming magnetic aspects and dynamics of the resulting quantum solutions involve novel phonon and roton driven (antisymmetrical) multi-spin induced orbital dynamics, subshell rehybridizations and valence shell rotations of central C and M atoms of these complexes by dense phonons, rotons and magnons within the quantum fluids (**known as the Little Effect[ 1]**), which determine important transformations, complexations, transport, orientation, organization, rotations and release dynamics of carbonaceous intermediates that determine the temperature (pressure, compositional and spin) gradient driven reaction trajectories during diamond formation by the various techniques of the direct, indirect and H-plasma CVD processes. Section 5 considers the application of this Resolution to the mantle formation of natural diamond. External magnetization organizes many domains of these intrinsic effects on macroscale for larger diamond formation. The use of external magnetic field is put forth here as a new useful technology for enhancing these intrinsic magnetic effects and ordering, organizing and correlating magnetic complexes over larger space in these older synthetic techniques for even faster, larger, and better diamond synthesis. Section 5 considers recent revelations of magnetized diamond synthesis. Section 6 determines magnetic instability of graphitic structures. Section 7 concludes this Resolution.

**2. The Cause of the Diamond Problem**

So why has it taken so long (150 years) to synthesize diamond? And why is it still difficult after 200 years to form large carat diamonds? And why is there a need for quantum consideration during diamond formation? Robert Linarres [2] recently noted: " the surface chemistry of how carbon atoms actually attach to the diamond lattice still remains murky." The Resolution here gives clarity to this murkiness for greater enhancing the understanding and improving the syntheses of diamond.

The cause and murkiness of the diamond problem are a result of the nature of the carbon atom: uniqueness and strength in covalent bonding; the varied options of covalent bonding; its low density of atomic electronic states; its electronic precise structures; its electric and magnetic hardness; its lack of inner p subshell; and its huge atomic instability. As a result, during bond rearrangement carbon atoms are electronically difficult to produce, control, alter and accumulate. From an older classical perspective, some prior investigators have sought to squeeze graphite to diamond based on diamond's greater density. Classically, bulk densities change continuously. But the strong covalent bonding and the required molecular and atomic scale electronic quantum dynamics require tremendous pressures and temperatures for such compression of graphite to diamond. The diamond forming pressures and temperatures are actually extreme enough to alter electronic densities of states. Such needed extreme pressure and temperature conditions need to disrupt electronically the strongest known electronic bosonic coupled covalent bonds over dimensions beyond atomic scales. Bundy [3] first achieved such extremely high pressures and temperature but such extreme conditions are impractical for large carat diamond production. On this basis, von Platen [4] imagined that no vice has the strength to compress graphite to diamond; he reasoned that it would break in the process. No extranuclear matter could sustain non-transiently the needed forces for such bulk compression to diamond. Also Bridgman [5] observed such difficulty of classically compressing graphite to diamond, thereby stating "graphite is nature's best spring."

The problem and murkiness of diamond formation are therefore a result of its formation being a quantum phenomenon not a classical dilemma. The electronic changes during graphite → diamond are nontrivial. Nonclassically, unlike bulk densities, electronic densities of states do not change continuously. During bond rearrangement of graphite to diamond, the changes in bosonic electronic energy and momenta are quantized and huge in magnitude and direction. During diamond synthesis conditions must focus huge energies tightly in space and time over many carbon atoms to alter electronics over states low in density. Among all elements, carbon atoms manifest the most number of these most difficult bond fixation events in forming diamond. Such bond rearrangements during condensations require bosonic rehybridization of s and p orbitals with diamond ($sp^3$) requiring greater sp orbital mixing than graphite ($sp^2$). Carbon atoms require more intense bosonic collisions and interactions to form diamond than graphite. Such complex necessary extreme collisions and interactions on the basis of quantum dynamics are the essence of the murkiness during C knitting into diamond. The probability of diamond formation is lower relative to graphite formation because such bosonic interactions must deliver specific quanta of energy and momenta for the sigma bonding of carbon to diamond. The magnitude of the required bosonic quanta of energy and momenta further diminishes the probability of diamond forming. Just the bosonic rehybridization energy alone for $sp^3$ formation is 401.9 kJ/mol [6]. The bosonic rehybridization energy per carbon atom is equivalent to hard-ultraviolet (UVC) photons. Such needed bosonic quantum dynamics require the accumulation of huge energies and momenta on bosonically coupled carbon atoms during the reaction trajectories to form diamond. **Not only does the total momenta and energy have to be huge, but the direction of motion (magnetic motion) is also quantized for specific momental constraints during diamond formation. Hence magnetic and roton quanta here are realized as important factors in this Resolution.** Therefore, the required bosonic electronic accelerations for fruitful diamond forming carbon bond rearrangement require nonclassical energetics, momenta and symmetrical dynamics for atomic fixation. The large activation barriers between graphite and diamond are hence bosonic quantum effects.

Such difficulty of atomic fixation is common among second row elements like carbon, nitrogen and oxygen [7]. The Haber process [8] exemplifies such fixation difficulties of forming all sigma bonds on nitrogen rather than pi bonds in the $N_2$ molecule. The difficulty and uniqueness of singlet oxygen relative to triplet oxygen exemplify fixation differences of oxygen terms [9]. Diamond fixation is more difficult because of the need to fix 4 hybrid carbon sigma bonds per carbon atom rather than 2 sigma bonds for O and 3 sigma bonds for N. Furthermore, in the case of diamond, C-C bonds must be fixed. The difficulty of electronic fixation results from the low density of electronic states and the lack of internal p symmetry (magnetic quanta) within the core first shell. Some elements of the 3d subshell and 4f subshell also exhibit fixation difficulty but to a more limited extent than the 2nd row elements (C, N, and O). The 3d and 4f subshells have higher electronic densities of states relative to 2p elements for less drastic fixation effects. Such higher densities of states of the 3d subshell in this case of diamond syntheses allow some 3d elements (Fe, Co, Ni) to provide a pattern (catalysis) (specific magnetic quanta) for diamond nucleation and growth.

In this case of diamond, before or during carbon atom fixation to $sp^3$ symmetry, the $sp^2$ bosonic symmetry must be broken. One of the causes of the diamond problem is breaking the $sp^2$ bosonic symmetry. Previous efforts have resorted to conditions to initially break $sp^2$ symmetry. Higher temperature, higher volumes and lower pressures determine conditions to break and overcome the strong and long range bosonic interactions associated with the $sp^2$ symmetry. The second cause of the diamond problem is the subsequent fixation of carbon atoms into $sp^3$ C bosonic symmetry. Such $sp^3$ fixation requires greater atomic and molecular interactions than $sp^2$ fixation. Such long range strong bosonic interacting intermediates and the difficult fixation contribute to the metastability regions between graphite and diamond in the phase diagram of carbon by Tammann [10]. $sp^2$ C does not require as much interactions and can form vibrationally on sub-picosecond time scales, involving only two carbon atoms (on the basis of gas phase formation of $C_2$ from $CH_4$ plasma). It is important to note that Steiner [11] determined that magnetization and consequent fermionic symmetry disrupt radical bosonic recombination on sub-picosecond time scales. **In this Resolution, such impact of electronic antisymmetry (fermionic) on ultrafast altering bonding kinetics is invoked later to show magnetic discrimination between graphite and diamond crystallization.** Therefore, $sp^3$ bosonic fixation to form diamond is accomplished by lower temperature, lower volume and higher pressure conditions. These two causes and their two contrary, compensating conditions present the murkiness and the dilemma for forming diamond and explain the large activation barrier between diamond and graphite. It seems necessary during synthesis to have both voluminous and cramped; both hot and cool; and both rarefied and compressed conditions, simultaneously. Prior researchers have partially overcome this dilemma of conditions via spatial gradients of temperature, density and pressure through which carbon atoms as various intermediates along various reaction trajectories form diamond. During the syntheses, one part of such gradients "cut off" or break the $sp^2$ bosonic bonding pattern and the other part of such gradients "cut on" the $sp^3$ bosonic bonding pattern. **Here in this Resolution, intrinsic magnetic aspects and consequent Fermi-Dirac statistics of such gradients are demonstrated as important for causing fruitful atomic and molecular dynamics for symmetry alteration for forming diamond. Moreover, external magnetization is demonstrated to induce faster organization of such diamond synthesis over larger space.**

These contrary conditions for breaking graphite and fixing C into diamond determine the diamond dilemma and create the large activation barrier between graphite and diamond and the smaller probability for pure carbon atoms to directly form diamond at ambient pressures in spite of the rather minute energy difference between the two allotropes. This diamond dilemma explains HT Hall's [12] observed large activation volume; Cannon's [13] demonstrated role of atomic processes during high pressure high temperature catalytic diamond formation; and Paul May's [14] realization that diamond formation essentially involves the atomization of graphite and the seemingly separate assembling of each carbon atoms into the forming diamond lattice. May's [14] atomization and assembly is here explained in terms of directly overcoming bosonic electronic interactions of carbon precursors (atomization) and in terms of directly rehybridizing the expanded state (assembly) into $sp^3$ carbon for diamond. As considered in the introduction, many scientists have explored this problem only to be frustrated by this dilemma. This dilemma contributes to the abundance of bulk graphite and rarity of bulk diamond on the earth's surface. This dilemma causes the monetary value of diamond by virtue of its rarity although its beauty. The future value of diamond will focus more on its properties and applications and its bulk formation by this complete solution. So a more complete resolution for larger diamond carats for future applications would be desirable. One resolution may be in the belly of the earth over long surface emplacement periods. But better solutions are needed due to current technological demands and the inability to pull large diamond carats from great depths beneath the earth's surface.

## 3. The Resolution of the Diamond Problem

Here it is suggested that a Resolution may easily be imagined if one could instantaneously "cut off" carbon-carbon bosonic interactions in pure carbon for decomposing the graphitic and carbynic precursors, thereby breaking the $sp^2$ and sp springs. Then, one could instantaneously "cut on" fruitful bosonic interactions for forming the $sp^3$ diamond spring. Furthermore, the impending pure $sp^3$ diamond spring must be protected from conversion to graphite as the pure carbon atoms relax and grow into the diamond lattice. It is a nuisance that the pure $sp^2$ graphitic spring readily forms from the nucleating, growing and relaxing $sp^3$ diamond spring under lower pressure conditions. But, beyond imagination, so how is it physically possible to do this with atoms and energy? In this work, a beautiful complete solution to this paradoxical

(classical) atmospheric dilemma for forming diamond involves the use of strong external magnetic fields via giant magnets for changing the electronic correlation (bosonic to fermionic statistics) between the radicals of the intermediary quantum fluids for the breakage and slowing of pi bonding symmetry to graphitic and carbynic structures and for the acceleration (via the Little Effect [1]) of sigma bonding to the $sp^3$ diamond symmetry. Here on the basis of quantum theory, it is suggested and demonstrated that the intrinsical and/or external strong magnetic field is a way to instantaneously break and change ("cut off") the bosonic sp and $sp^2$ interactions and to create quantum fluidic carbon intermediates by changing the correlation of electrons on the activated atoms of the radical intermediates, thereby cutting off or reducing paired electronic covalently bosonic interactions under synthetic conditions. The magnetization and consequent fermionic media therefore create high spin $sp^2$, metastably prevent $sp^2$ regraphitization, and prevent $sp^2$ relaxation to the ground state electronic configuration of carbon and subsequently transform $sp^2$ to $sp^3$ symmetry! The fermionic antisymmetry prevents bosonic collapse to undesirables. The magnetization and consequent antisymmetry of the fermions slow relaxation thereby allowing the efficient conversion, organization and accumulation of thermal energy for feasible adiabatic orbital orientation, subshell rehybridization and shell rotation. Furthermore, here it is demonstrated that the intrinsical and/or external magnetization orient electrons of multi-radicals in the resulting intermediates for phonon-roton driven multi-spin induced spin orientation, orbital orientations, subshell rehybridizations and shell rotations (the Little Effect) for ("cut on") $sp^3$ fixation across temperature, compositional, pressure and spin gradients that determine the reaction trajectories from hotter to cooler regions of the gradients that transform $sp^2$ carbon to $sp^3$ carbon, protect $sp^3$ carbon, orient $sp^3$ carbon and release $sp^3$ bosonic carbon to deposit diamond at a growth edge. Furthermore, the magnetization rotates carbon radicals, molecules and nanodiamond for organized, orchestrated consolidation into the growth edge! Here the Resolution determines that external magnets enhance these intrinsic magnetic effects over bulkier spaces.

It is important to note that the characteristics of the atoms of the elements associated with diamond formation are well suited for such a fruitful magnetic enhancement (cut off $sp^2$ and cut on $sp^3$) of diamond synthesis. The carbon atom is suitable and capable by its electronic precise atomic structure in some states and terms to exist with 4 lone electrons in the same shell for multiradical atoms of very strong magnetic moments per atom. The resulting 4 frontier radical orbitals individually exhibit very hard directional aspects of motion for strong orbital magnetism. C atoms, diamond embryos and nanodiamond intermediates can exist with varying broken $sp^3$ C-C bonds for high spin and magnetism. T. Enoki [15] recently demonstrated such high spin and magnetism in synthetic diamond frozen in the product during dynamical synthesis by explosion. The catalysts (Fe, Co, and Ni) also exhibit high spin and high magnetic moments per atom for favorable coupling to these magnetic carbonaceous intermediates. By Hund's Rule, the half filled $2sp^3$ and 3d subshells energetically favor high spin magnetic electronic configurations of C, Fe, Co and Ni atoms and aggregates in the diamond forming quantum fluids. It is important to note here that this suitability of carbon atoms and metal catalysts for magnetized diamond formation are substantiated by recent controversy surrounding magnetism in carbon [16]. Although the magnetic and high spin influence on carbon crystallization do not depend on the static ferromagnetism of carbon, this initial magnetic revelation of discriminating diamond and graphite formations inspired such other researchers to study magnetism in carbon allotropes. The many recent findings of magnetism in carbon, carbon-ferrometal alloys and hydrogeneous diamond provide strong evidence for this Resolution.

Therefore once melted and antisymmetrically oriented under strong magnetization, the carbon, metal, and hydrogen atoms are slower in relaxation by rebonding, especially rebonding to pi and aromatic bonds. The change in spin statistics slows the recrystallization for greater control of intermediate states. Thereby the activation energy for graphitic decomposition is lowered. From quantum theory, in this Resolution, antisymmetry (thought to hold-up the neutron stars from collapse to blackhole) slows and prevents graphitic collapse of the intermediary high spin $sp^3$ carbon radicals in the intermediary quantum fluids. The antisymmetry also prevents solidification of the catalysts [17] for liquid catalytic diamond growth at lower temperatures. Such kinetic effects of antisymmetry on decelerating pi, conjugated and aromatic bondings and decelerating relaxation to the ground carbon electronic state are in analog to antisymmetrical slowing of phosphorescence relative to fluorescence [18]. Magnetization traps the atoms metastably in $sp^2$ and $sp^3$ hybrid states in the quantum fluids. Furthermore, the magnetization and the resulting unpaired electrons of the atomic, fluidic states by antisymmetry change the permittivity and permeability of the media for weaker carbon interactions for ease of breaking the carbon precursors into these quantum fluids on larger spatial and smaller temporal scales with the consequent lesser needed expansion (mechanical volume) and greater $sp^3$ carbon radicals accumulation and stabilization for supersaturation of antisymmetrical $sp^3$ carbon for greater probability to diamond and larger single crystal formation.

      The growth conditions during the various processes (such as this graphitic breakage and spin orientation) determine gradients of temperature, pressure, density, composition, spin and magnetism. The hotter parts of such gradients decompose and atomize precursors, forming the magnetic quantum fluids. **It is important to note that the energy of the gradient for such atomization does not come from the magnetic field. Ovens, lasers, microwaves, hotwires, ect. supply the bulk of the activation energy across the gradients.** The intrinsic magnetization only orients resulting electronic spins on activated C, M, and H atoms and between activated C, M, and H atoms of the quantum fluids, creating and stabilizing the resulting fermionic electronic states and radicals in the fluids. The characteristic synthesis conditions during the HPHT direct, the catalytic HPHT indirect, and the metastable CVD processes cause growth zones with these suitable gradients. See Figure 1. These gradients in these zones drive the bond breakage, spin orientation, subshell rehybridization, radical complexation, shell rotational dynamics, orientation of carbonaceous complexes, molecules and nanodiamond, and electronic spin pairing for diamond nucleation and growth as described by characteristic reaction trajectories, involving the relaxation and conversion of these quantum fluids to diamond. Such complex processes during diamond nucleation and growth are the essence of the murkiness. The reaction trajectories are driven by these temperature, pressure, density, compositional, spin and magnetic gradients across the growth interfaces. Intrinsic magnetism is induced in these gradients of the growth zones.

      The resulting magnetized quantum fluids provide spin antisymmetry for the breakage of $sp^2$ bonds via the lone electron complexation of graphitic antibonding molecular orbitals by radicals of these quantum fluids. The high spin antisymmetric rapidly rotating radical atoms in the hotter part of the gradients catalyze ¶ bond cleavage for graphitic decomposition. Such radical rotation and consequent spiral of electrons in and out of antibonding orbitals of the graphitic structures induce magnetic flux in the graphitic structures for magneto-catalytic breakage of ¶ bonds. The rotating radical solvent media of the quantum fluid induce magnetism in the graphitic precursors for magnetized symmetry breakage of bosonic bonds to form more fermions. The rotational breakage of ¶ bond and aromaticity is greater because of the more compatible coupling of roton motion to the ring currents and ¶ bonds relative to the densely localized $sp^3$-$sp^3$ C-C bonds. However, in the hottest regions of the gradient across the growth zones both ¶ and σ bosonic bonds are broken due to the faster fermion rotational energies. The faster radical rotational energy in the hotter regions provides energy to better couple to the σ bosonic covalent bonds for their symmetry breakage and decomposition to fermionic radicals. As the media cools across the growth zone, the σ bosonic covalent symmetry crystallizes first with ¶ bosonic symmetry, requiring lower temperatures. The catalytic complexation however locks C atoms into $sp^3$ sigma bonds of diamond before cooling to graphitization.

      The conjugate pi bonds and aromatic bonds of the graphitic precursors are superconductive on the molecular scales. The lone electrons of the surrounding radicals in the quantum fluids are magnetic. Such magnetic lone electrons and superconducting ring currents

within the same correlated graphitic structures do not mix, thereby catalytically driving the graphitic instability and its decomposition. The strong magnetization, ligating radicals, and thermal energy thereby break the molecular superconductivity of such conjugate pi and aromatic bonds in the graphitic structures for their ready transformations to radical fluids and their subsequent crystallization to the more stabilizing diamond lattice. Therefore, the magnetization and the consequent high spin quantum fluids interact unfavorably with the delocalized electronic motion of ring currents in graphitic structures, thereby causing graphitic instability under synthesis conditions for the graphitic decomposition driven by a Meissner effect [19] on the molecular scale. The radical quantum fluids thereby drive and catalyze graphitic decomposition, forming radical quantum media.

**G. N. Lewis [20] first determined that the external magnetic field can be used to study such individual radicals.** But here for the first time R. B. Little [1] considers the novel bonding mechanics, orbital symmetry transformations and electronic fixation in such antisymmetric, dense fermionic, high spin, magnetic quantum fluids. The gradients and antisymmetry provide novel environments for new chemical dynamics. Here, R. B. Little thereby demonstrates that very strong magnetic field can organize bond rearrangement within such dense radical environments for symmetry breakage of Lewis covalent bonds via the dense radical intermediates in these quantum fluids. On this basis, novel high spin fermionic conditions are determined for the frustration of the Woodward-Hoffman Rule [21]. Furthermore, here it is demonstrated that the magnetization lowers the needed thermal activation for breaking pi bonds and breaking $sp^2$ symmetry to form $sp^3$ bonding symmetry. Here it is demonstrated that such magnetization influences chemical dynamics and does not require external photo-excitation as in the El-Sayed Effect [22]. However, the bond rearrangement and atomic fixation occur adiabatically by the Little Effect[1].

Whereas on the basis of **Kasha's Effect [23], phonons in bosonically coupled atoms in molecules are readily produced by the nonadiabatic excitation of upper level electronic states. But, here on the basis of the Little Effect [1], the fermionic coupled radicals in these magnetic quantum fluids orient spins so that antisymmetry slows relaxation by phonon scatter and release.** Whereas bosonically coupled electrons by the Kasha Effect [22] readily dissipate electronic potential via vibrational modes, fermionically correlated electrons have much weaker vibrational energetics for less efficient dissipation of electronic potential energy vibrationally and stochastically. The fermions allow efficient roton distribution of energy. **On the basis of the Little Effect, the antisymmetric electrons of the multi-radical atoms readily rotate and spiral for twirl of valence orbitals, subshells and shells, each of which containing lone electrons for the ready conversion and interchange of electronic potential energy with rotational and magnetic energies of shell, subshell and orbital. Such electronic relaxation via rotational energy modes of fermions is more efficient than electronic relaxation via vibrational modes of bosons. This leads to efficient roton correlated radicals for energetic accumulation to form diamond. For such rotational motions, the electronic orbital moments orient with the spin moments for the (nonstochastic) organized rotational kinetic energy and its ready accumulation and conversion to electronic potential energy for causing electronic excitation or also assisting electronic relaxation into various orbital motions and rehybridizations.** Unlike bosonically coupled electrons and their difficult vibrational accumulation of energy for electronic rehybridization and shell and subshell transitions, fermionically coupled electrons via roton and phonon quanta and motions more easily order their kinetic energies rotationally for energetic accumulation and organization for easier spin-rotational adiabatic orbital dynamics for rehybridization.

Thereby, the magnetization allows the accumulation and concentration of organized vibronic, rotational and magnetic energies within the correlated, dense fermions for faster, bulk, asymmetric, adiabatic rehybridization dynamics of fermionic carbon atoms: $sp^2 \rightarrow sp^3$. By the Little Effect [1], the resulting concentrated optical phonons-rotons drive bulk multi-spin induced orbital revolutions, subshell rehybridization and shell rotation about fermionic atoms in the quantum fluids. See Figure 2. Therefore, in addition to breaking the precursors by the magnetization reducing C---C bosonic interactions, here it is considered that magnetization allows high spin induced transformations of orbitals for spin oriented orbital symmetry breakage for $sp^2 \rightarrow sp^3$ fermionic rehybridization by the high antisymmetrical spin density alteration of orbital mechanics. These chemical phenomena of the Little Effect are consistent with physical effects of far from equilibrium symmetry breakage of Prigogine [24] and the dynamical symmetry breakage by magnetic field of Miranski [25]. On the basis of such antisymmetrical accumulation and organization of roton energy and fermions, no external photons are needed to excite rehybridization mechanics. **The Little Effect [1] is therefore different from the El-Sayed Effect [22] (that photo-excited orbital states can couple strongly with electron spins to cause intersystem crossing nonadiabatically), but here on the basis of the Little Effect the dense phonons, rotons and magnons in the quantum fluids are determined to affect orbital dynamics adiabatically**, via the lone electrons of complexation binding antibonding $sp^2$ orbitals of the graphitic carbon, causing spin interaction with the ring current within the graphitic structures for destabilizing and lowering the bond order of the graphitic structures. Such catalytic activated pi bond cleavage is consistent with electric charge activated diamond nucleation (biased enhanced nucleation) of Yugo [26]. Negatively charging and polarizing carbon atoms or carbanion formation [26] also facilitate the breakage of bosonic bonding of graphitic and carbyne precursors, just as the radicals serve the role of electrons in occupying antibonding orbitals for graphitic decomposition during biased enhanced nucleation BEN. The high spin magnetic quantum fluids provide a conducive environment for sufficient chemical potential energy (to replace the high temperatures), low permittivity and high permeability (to replace the high voluminous conditions) and suitable chemical pressure (to replace the high mechanical pressures) for lowering quantum trajectories to forming diamond.

The $sp^2$ carbon centers in the resulting quantum fluids are less symmetrical than $sp^3$ carbon centers so that the $sp^2$ carbon centers are more subject in hotter regions of the gradients to roton driven multi-spin torque of valence electrons (the Little Effect) into $sp^3$ hybrid states. See Figure 2. The resulting $sp^3$ carbonaceous quantum fluids are more stable in the dense radical and strong magnetic environments relative to the lesser stability of $sp^2$ radical and graphitic structures in the radical and magnetic environments. The ring currents in graphitic structures are incompatible with the surrounding media of radicals and the strong magnetization on the basis of the Meissner Effect [19].

On the basis of the decomposition of graphitic precursors and the rehybridization of $sp^2$ C to $sp^3$ C, the resulting $sp^3$ fermionic carbon atoms in time cool and undergo ordering via interactions. The interaction between the resulting $sp^3$ carbon fermions is more of a van der Waals and magnetic interactions, which contribute to magnetic ordering for the possibility of a liquid crystalline ferro (a magnetic quantum liquid) assembly of diamond under appropriate cooler conditions across temperature, pressure, compositional and spin gradients in the quantum fluids. See Figure 1. Many investigators have observed evidence of this external orienting of radical spins, self orientation of lattice defects, and quenching defects within the diamond lattice. In such a high spin magnetic quantum liquid state, the carbon radical species interact with neighboring radicals antisymmetrically via individual oriented lone unpaired electrons in bonding and antibonding orbitals of neighboring complexes, such that the complexation (($CC_4 \cdot C_x$), ($CC_4 \cdot HM_x$) and ($CC_4 \cdot H_x$)) of the $sp^3$ C species via its nonbonding molecular orbitals, causes zero or lower bond order within fluidic phases. See Figure 3. It is important to note that across the gradient from hotter to cooler, the spin induced swirl and twirl of electrons, orbitals, subshells and shells disrupt and interchange bonding and antibonding molecular orbitals and interactions in the hotter zones with slower spiral dynamics in cooler zones for slower swirl and twirl of orbital, subshells and shells for locking atoms into bonding molecular orbitals and displacing antibonding ligands. Other researchers have observed the creation, existence and origin of remnants of these clusters in the final diamond product [27]. These radicals of antisymmetrical complexation (($CC_4 \cdot C_x$), ($CC_4 \cdot HM_x$) and ($CC_4 \cdot H_x$)) in the quantum fluids physically attract and polarize their high spins for ferromagnetic or superparamagnetic quantum fluid solutions along the reaction trajectories. Researchers observed the orientation of magnetic centers in diamond [28]. These radicals form complexes: (($CC_4 \cdot C_x$), ($CC_4 \cdot HM_x$) and ($CC_4 \cdot H_x$)). The complexes interact magnetically. The experimental work of other investigators prove such magnetic properties of the complexes [29] their spin exchange interaction [30] via the diamond lattice, their orientation by the external magnetic field and their ferromagnetism at high defect concentrations. Recently, Peng and coworkers [31] observed magnetism of high concentrations of NiMnCoC complexes in synthetic diamond formed at pressures too low for sufficient crystallization. Peng et al also observed intrinsic magnetism in fracture surfaces of black diamond containing many cavities but no inclusion [31]. As put forth in this Resolution, such magnetization of the fluid forming media and its residual impurities magnetically couple with radical states and dangling bonds thereby preventing their graphitization and protecting the diamond as it forms. The magnitude of the magnetic interactions varies with the nature of ligating fluids with the Fe media having the strongest magnetic interaction and ordering. The Fe quantum liquid affords stronger antisymmetrical prevention of graphitization and faster rehybridization of $sp^2$ to $sp^3$ for faster, larger diamond growth. This is consistent with experimental observations of fewer Fe complexes in diamond relative to the Co, Ni, and H defects. The Fe complexes anneal better due to greater moments. The conditions induce rotation of ligands of clusters and also rotation of valence shell of central atoms of clusters by the Little Effect [1] across the gradients.

Furthermore, this spin orientation among atoms of the diamond forming quantum fluids allows the accumulation of a multitude of high spin $sp^3$ carbon atoms with electronic orbital and spin correlation, orientation and organization for diamond crystallization as the growth zone cools. See Figure 1. As the antisymmetrical $sp^3$ carbon fermionic quantum fluid cool the rotons and magnetism transform to spin paired bosons with vibronic release of energy. The rapid cooling at the interface quenches and locks the carbon atoms into the $sp^3$ diamond lattice. Other researchers have demonstrated these dynamics reversibly by imposing synthetic conditions on diamond to create new magnetic impurities and anneal prior existing magnetic impurities in the diamond lattice [32]. Researchers have observed that electron, proton and neutron irradiations of diamond alter magnetic impurity complexes in the lattice [33]. This use of external fermions and their creation of magnetic lattice centers in diamond is direct evidence of this Resolution. Because the dynamics of these carbonaceous complexes in such quantum fluids in the hotter regions are dictated by antisymmetrical spin, motion and changing internal magnetic field, such liquid, crystalline, quantum catalytic fluids allow the multispin interactions within and between these magnetic complexes in hotter parts of the gradient for dense vibron-roton-magnon induced spin orientation, subshell rehybridization, orbital dynamics and novel valence shell rotation of radicals of the complexes across the gradients on the basis of the Little Effect [1]. See Figure 1. Such high spin environments of H plasma and metal quantum fluids also result in rotational-magnonic torque of valence shell orbitals for the ready rotation of bonding and antibonding molecular orbitals between neighboring (ligating) radicals and complexes (See Figure 4) of atoms in the hotter parts of the gradient with slowing of rotation in cooler parts of gradients for chemically locking bonds into $sp^3$ C-C in such an ordered quantum fluids toward cooler zones across the growth interface. See Figure 1. As the media cools, the rotational motion slows and C-C bonds get locked into place. Indeed, the chemistry of adding C to diamond is murky but this magnetic quantum Resolution gives clarity to the mystery. Different liquid radicals have different magnetic moments for different rotational and rehybridization dynamics and different catalytic activities to form bigger diamonds: M>H>C.

In addition to the Resolution accounting for the magnetization and its role on the direct process of forming diamond from pure carbon, the Resolution also accounts for the metal quantum fluids and H quantum fluids of the indirect processes and the plasma CVD processes, respectively.  The metal atoms and H plasma radical-fluids can serve the same role as carbon radicals under less stringent conditions for forming diamond.  Actually the observed, induced and sustained ferromagnetism in the catalyst under prevailing conditions of the Fe, Co and Ni catalytic formation processes is proof of the here discovered magnetic quantum Resolution.  Furthermore, the observed enhanced diamond formation by electric current, DC biasing the substrate and weakly magnetizing plasma during the CVD methods for prior subtle but unappreciated magnetic effects is further proof of the use of the magnetic quantum Resolution for the metastable processes.

       This quantum fluidic state was originally put forth here for novel chemistry by R. B. Little [1], but Bigelow [34] subsequently in "Spins Mixed Up", independently reviews similar spin and collisional aspects of atomic radicals for the physical consideration of Bose Einstein Condensation among atomic radicals.  Bigelow's Bose-Einstein Condensations stably exist in the earth's weak magnetic field, but the imposed strong magnetization of R. B. Little results in fermionic ferromagnetic or superparamagnetic quantum fluids rather than the BEC quantum fluids of Bigelow.  Such ferromagnetic or superparamagnetic ordering of carbon atoms in these quantum fluids may be fleeting or transient but such ordering persists along diamond forming reaction trajectories for substantial novel catalytic (by the Little Effect) [1] effects in the Fe, Co, and Ni quantum fluids or highly compressed carbon quantum fluid or highly thermal hydrogenous quantum fluid at diamond surfaces.  Whereas the spin ordering in Bigelow's systems results from statistics of Bose-Einstein, the chemical dynamics of the Little Effect involves Fermi-Dirac statistics.  It is important to note that high pressures increase spin coherence and correlation as suggested here in agreement with Bigelow [34].  However under Bigelow's conditions, the collisions maintain BEC states.  Whereas on the basis of the Little Effect, the collisions at higher pressures sustain fermionic exchange for magnetism.  It is important on the basis of the Bigelow and Little to give clarity to a misconception of thermal effects and spin correlation.  Higher temperatures merely diminish exchange effects between spins, making it easier for external alteration of spins.  As noted by Bigelow for such gaseous conditions momentum is still conserved between collisions for adiabatic processes.  Therefore in these quantum fluids as temperature is raised the effect of the spin exchange force is diminished but for a spin ordered state the order is maintained due to conservation of spin momentum.  Use of external magnetic field polarizes the spins over larger space such that despite high temperatures, conservation of momentum maintains spin order beyond the Curie temperature for novel effects.  On this basis of external magnetization, the Little Effect determines novel catalytic phenomena to complement the Hedvall Effect [35].  Whereas high pressure increases spin exchange, high temperature decreases spin exchange effects.  So by increasing both temperature and pressure the exchange is maintained for persisting ferromagnetism even at extreme conditions.  HPHT conditions during diamond formation cause sufficient interactions to induce and sustain ferromagnetism in carbon quantum fluid of the direct and the metal quantum fluid of the indirect processes under diamond growth conditions.  Thereby the extreme HPHT diamond forming conditions induce intrinsic magnetism which by this Resolution organizes by Fermi-Dirac statistic the diamond formation.  Observations of other researchers provide evidence of this here suggested lattice pressure induced ferromagnetism of defects and external high pressure high temperature induced ferromagnetism of high concentrations of defects and interstial atoms [36,37].  In addition, the high temperature plasma causes sufficient, strong interactions between H atoms for local magnetic organization.  In addition to carbon and metal atoms, other catalytic elements may exist in this state with carbon under more severe conditions for diamond formation.

       Therefore on the basis of the Little Effect, the resulting intrinsic magnetization of the quantum fluids in the direct, indirect and H plasma processes and the consequent high spin and motion driven dynamics across the gradients of the growth zones cause spin barriers (frustration and asymmetric transformation) to Woodward-Hoffman [21] like (symmetry driven) orbital dynamics.  The Woodward-Hoffman Rule determines that the broken carbon bonds would

preserve their symmetry and undergo related faster kinetic of re-graphitizing by symmetrically compatible and already existing $sp^2$ orbitals to reform the graphitic precursor rather than the slower pathways of rehybridization for bond rearrangement for forming different $sp^3$ orbitals of different symmetry for diamond formation.  But, here the Resolution demonstrates a high spin asymmetric bottleneck for radical carbon, ferrometal and hydrogen radicals by the intrinsic magnetism and the enhancement by external magnetization for disrupting orbital symmetry during bond rearrangement, thereby preventing graphitization, enhancing transformation of graphitic $sp^2$ carbon to $sp^3$ carbon, and substantially lowering the $sp^3$ bosonic bonding interactions between carbon atoms so the $sp^3$ carbon atoms can accumulate and densify for greater probability of bulk (macro) high spin $sp^3$ carbon radical interactions for larger diamond formation.  The novel spin, magnetism and motion on the basis of the Little Effect [1] frustrate the Woodward-Hoffman Rule under very different extreme conditions, causing orbital symmetry breakage for $sp^2$ to $sp^3$ rehybridization.  Woodward-Hoffman applies to different statistics of Bose-Einstein, but the Little Effect and the broken symmetry is a consequence of Fermi-Dirac statistics and its impact on bond rearrangement dynamics.  Therefore, Little [1] contributes fermionic radicals in quantum fluids for the novel additional spin scenarios during the processes of the direct, indirect and H plasma quantum fluids for explaining orbital revolutions and rehybridizational dynamics by the Little Effect during chemical reactions under extreme reaction conditions that are much different from Fukui-Hoffman [38] orbital symmetry and dynamics.  The importance of orbital symmetry for pathways of chemical reactions has been demonstrated and led to the Nobel Prize to Hoffman and Fukui on the basis of the Woodward-Hoffman Rule.  Here, the Little Effect expands the knowledge of chemical dynamics by providing a new avenue of chemical synthesis and understanding chemical properties.

During the last twenty years, many have contributed new experimental verification of electronic spin effects on photophysics [39,40] on photochemistry [41,42], on radical recombination [43,44], on spin chemistry [45], on spin catalysis [46,47] and on spin isotope effects [48].  The Little Effect is distinct from these prior spin phenomena.  The Little Effect is very different from the radical pair effect.  The Little Effect differs because it involves interactions of 3 or more multi-radicals, and their motion and collisional interactions for consequent multi-spin, roton and phonon induced valence shell orbital rotational and orbital rehybridizational dynamics.  The radical pair effect considers just the dynamics between two radicals.  Whereas the spin chemistry of Turro [43], Buchachenko [45], and Hayashi [44] are more limited to spin itself, the Little Effect differs in its determination of spin alteration of orbital subshell and shell dynamics.  The Little Effect is multi-radical phenomena and it determines novel orbital, subshell, shell and even atomic and molecular dynamics due to dense spins and energy.  In Resolving the Diamond Problem, Little makes use of both orbital and spin nonclassical mechanics on a larger spatial frame in a shorter temporal range.  Little combines these two electronic aspects of spin and orbital motions of chemical reaction dynamics to supplement the Woodward-Hoffman Effect [21] and to advance chemical dynamics in a general fashion, thereby applying spin-orbital mechanics to resolving the most difficult problem of concerted, correlated multi C-C single bond formation for diamond formation and synthesis.

On the basis of Fukui and Hoffman, reaction pathways involving orbitals of similar symmetry are more probable.  The transformation of $sp^2$ to $sp^3$ requires change in symmetry with the consequent lower probability.  R. B. Little employs, controls and manipulates spin states by strong external magnetization to slow bonding of existing or rapidly forming $sp^2$ orbitals to allow more time for such symmetrical orbital transitions for the rehybridization to the $sp^3$ state and accelerated $sp^2$ to $sp^3$ rehybridization over larger space.  Beyond these intrinsic magnetic effects of smaller domains associated with polycrystalline diamond formation, R. B. Little uses external magnetic force to heighten and to better control spin states; to enhance the polarization of electronic spins of radical intermediates over macrospace; to better reduce hybridized relaxation to ground atomic electronic states; to thereby better slow $sp^2$ bonding; to better accelerate spinor dynamics of rehybridizing $sp^2$ carbon to $sp^3$ carbon; to better protect $sp^3$ carbon via spinor ligands in the quantum fluid; and to better accumulate nonbonding high spin $sp^3$ carbon and catalyst atoms over multidomains over macrospace until feasible for more enhanced spin inversion by multi-spin shell rotations for $sp^3$ bosonic pairing of electrons and bonding of carbon atoms to diamond over macrospace all on the basis of the Little Effect [1]. The extreme external magnetization enhances fruitful dynamics of the intrinsically magnetic quantum fluid for producing diamond. R. B. Little uses external magnetic field to manipulate spin dynamics to change and to control orbital dynamics by the Little Effect for greater probability of bonding to diamond in shorter time and over larger spatial dimension. Whereas intrinsic magnetism may be limited to micro-domain sizes, external magnetization organizes multi domains for larger faster size crystal diamond formation. **It is important to note that the external magnetic field steers activated energized radicals for organizing and orchestrating their relaxation diamond formation by the Little Effect [1], the magnetic field does not power diamond formation. The magnetic field does not activate atoms. The bulk of the power and activation energy come from an oven, electric current, furnace, microwave or laser. The magnetic field just modifies activated carbonaceous, ferrometal and/or hydrogenous high spin intermediates formed by these power sources. So these intermediates cannot due to antisymmetry relax to graphite but instead organize and relax to diamond.** Just as the nonclassical nature of electrons and electromagnetic interactions prevent the collapse of the electron to the nucleus with the resulting atomic structure, so also does the wave nature of fermionic electrons during chemical transformation and the use of appropriate external electro-magnetic fields prevent collapse of $sp^3$ diamond intermediates to graphitic structures during the magnetic carbon allotropic conversion to diamond.

**4. Natural Diamond Formation**

In support of this Resolution, it is important to keep in mind that Kimberlite rock pipes where diamond is mined from the earth have unique magnetic properties relative to surrounding rock regions [49]. Such magnetic distinction has led to the location of diamond bearing Kimberlite based on airborne measurements of variations in the earth's magnetic field on the basis of so called magnetic anomalies [50]. Here this Resolving model on the basis of the Little Effect is consistent with the natural formation of diamond in the earth's mantle. Natural diamond formed within the mantle of the earth billions of years ago under the conducive high pressures and high temperatures and possibly iron catalysts within the earth's mantle. Nature overcomes the diamond dilemma within the mantle by the high pressure, high temperature and catalytic activity of Fe on carbon dioxide and water or other carbonaceous sources within the earth's mantle. The mantle's conditions produce the magnetic quantum fluids of Fe, C, H, and O in consistency with this Resolution. Diamond crystallizes from such quantum magma in the earth's mantle. The human synthesis of diamond is somewhat a result of mimicking mantle conditions on 1.) pure carbon for the direct method or 2.) carbon and ferrometals for the indirect method. Mantle diamond is brought to the surface of the earth by volcanic emplacement within fluidic rock channels called Kimberlites and related Lamproites. Such volcanic emplacements involve fractures in continental cratons and are very rare [51]. The resulting release of Kimberlitic magmatic fluids comes from depths over 100 km. Kimberlite is volcanic residual magmatic fluid containing high amounts of iron, carbon dioxide, water and potassium. However upward emplacement in the hot fluidic Kimberlite to the lower pressures at the earth's surface leads to diamond instability and its graphitization within the hot magmatic rock-fluid. On the basis of the Resolution here, R.B. Little demonstrates how magnetic properties of the iron magmatic fluids in the mantle organize diamond formation and protect diamond during its emplacement from the earth's mantle to the surface of the earth. Galimov [52] determines isotopic effects in Kimberlite magmatism that provides support of this magnetic Resolution of natural diamond formation. It is important to note that under mantle conditions, the high pressures and high temperatures sustain the ferromagnetism of the magmatic fluids! The emplacement is thought to involve explosive volcanic activity wherein Kimberlitic magmatic fluid from depths of 100 kilometers burst upward through the earth's crust toward the surface of the earth. The rapid ascent speed is thought to be 10-30 kph [53]. Here the Resolution suggests that the magnetic properties of the magmatic fluid assist in protecting the mantle diamond in the mantle and during its emplacement to the surface of the earth. Below it is considered that magnetism protects diamond abrasives during grinding. Therefore, the key ingredients to natural diamond formation include high pressure, high temperature and catalysts and, as put forth in this Resolution, magnetism. By developing technologies to mimick mantle conditions man has given partial solution to the diamond problem. Some partial success and solution to this dilemma have involved the use of liquid metals of group VIII and the use of atomic hydrogen for the indirect and H plasma. The Resolution here gives complete solution by external magnetization.

**5. Magnetized Diamond Formation**

These common roles of magnetic and orbital-spin phenomena during diamond formation in the direct autocatalytic, in the indirect catalytic, in the H assisted vapor deposition and in the earth's mantle processes resolve the diamond problem. These common roles further suggest a complete solution by use of external magnetic fields for controlling the intrinsic magnetic and orbital-spin dynamics for the complete Resolution of the Diamond Problem. Here is some evidence of what is to come by using external magnetism to ushers in a new era in diamond synthesis. Little [54] first discovered the accelerated formation of diamond in strong magnetic field (>15 Tesla) at atmospheric pressure by using Fe catalyst and carbon precursors. Wen [55] has subsequently observed similar effects of magnetic field on diamond formation although under the expansion pressure of sealed hot Fe pipes. Recently, Huang [56] suggest using magnetic field and microwave plasma on carbon polymer in order to enlarge a diamond seed. Druzhinin et al [57] used ultrastrong magnetic field (300 Tesla) for diamagnetic compression to crystallize diamond from graphite, but erroneously assumed the magnetic field only caused compression of graphite for diamond crystallization. Druzhinin did not explicitly use external magnetic field to directly influence the chemical dynamics. Druzhinin et al [57] did however note that the magnetic compression resulted in faster larger diamond crystals than comparable compression by traditional mechanical high pressure high temperature techniques. Druzhinin did not entertain, as done here in this Resolution, that the magnetic field itself contributes additional beneficial catalytic effects via radical intermediates for facilitating diamond nucleation and growth. Much weaker magnetic fields (> 1 Tesla) have been employed for over a decade to focus plasma during CVD; but not as put forth here wherein strong magnetization affects the antisymmetrical bond rearrangement at the growth interface as by the Little Effect [1]. Hiraki [58] used magnetic field in microwave plasma CVD but fields of much weaker strength than RB Little, serving simply to spread and densify the plasma for more uniform deposition by the Matsumoto style synthesis. Also, Wei [59] used that external magnetization to allow better internal coating of high aspect ratio tubes with diamond like carbon and SiC. R. B. Little is the first to predict, observe and discover that strong magnetization causes intrinsic nonclassical dynamics for enhancing bond rearrangement to diamond. Recently, Skvortsov [60] observed diamond formation in the strong magnetic monofields created by lasers.

In addition to supportive data of this Resolution by magnetic diamond formation, researchers have also observed that magnetization slows diamond decompositions during abrasion and grinding. Kuppuswamy [61,62] observed that external magnetic field affects and slows electrochemical etching and grinding of cemented carbides, diamond, SiC and $Al_2O_3$. Other researchers have reported the effect of magnetization in slowing decomposition of ferroabrasives [63]. Magnetism enhances metal removal but slows the damage to the diamond used to grind the metal. At only, 50-100 Gauss on 15% Na $NO_3$ removal rates of tungsten carbide was greater under the external magnetic field. On the basis of this Resolution, the magnetism slows the decomposition chemistry of diamond under the local high temperature and high pressure created by grinding processes such that graphitization is slowed and the diamond is protected. External magnetic field slows pi bonding, which is crucial for abrading diamond. ¶ bonds form due to impact and resulting high pressures and high temperatures during grinding but under external magnetic field and friction induced HPHT the diamond surface will not graphitize. The surface radicals formed during abrasion resist graphitization under the antisymmetry caused by the external magnetism. Graphitic structures are unstable in the radicals and magnetism.

**6. Magnetized Graphitic Instability**

Experimental data of other investigators supports this graphitic instability in dense radical environments and under strong magnetic field. The diamond lattice is more stable than the graphitic lattice in the radical and magnetic environments due to the weaker coupling between the more localized dense $sp^3$ C-C bonds and lattice radicals. On the other hand, there is stronger unstable coupling between the more delocalized graphitic ring currents and formed lattice radicals. In support of this spinophillicity of diamond, ferromagnetic states in densely defective diamond have been observed [64]. Partridge et al observed ferromagnetic transfer into outer diamond coating on ferromagnetic cores when ferromagnetic metal rods are coated with diamond [65]. Putov et al. [66] observed that thermomagnetic treatment of steel produced excellent steel. There is a lot of evidence that the magnetization and high spin radical environments favor the spinophillic diamond lattice over the spinophobic graphitic lattice. Researchers have demonstrated the greater stability of spin in the diamond lattice and the formation of carbon onions in magnetic field by thermalizing nanodiamond in external magnetic field [67]. Whereas other researchers have demonstrated the weaker stability of spin in the graphitic lattice, by thermalizing nanodiamond in zero applied magnetic field to form turbostratic graphite and nanographite [68,69]. R. B. Little [54] observed that thermomagnetization (> 15 Tesla) of nanoparticles of Mo-Fe with flowing $CH_4$ and $H_2$ leads to nucleation of diamond at atmospheric pressure rather than the formation of CNT under similar conditions but in zero applied magnetic conditions. Yokomichi [70] observed this instability of graphitic structures and strong magnetism by the collapsed of growing CNT in strong external magnetic fields of 10 Tesla. B Wen [71] observed what he calls new diamond by thermomagnetized (10 Tesla) catalytic transformation of carbon black under higher pressure and Fe nanocatalysts to form this intermediary new diamond, thought to be intermediary between rhombohedral graphite and diamond. It is important to note that B. Wen et al.[55] needed background thermal expansion pressure at 10 Tesla to form diamond otherwise at 10 Tesla and atmospheric pressure they would have gotten the collapsed CNT of Yokomichi et al [70]. Therefore in the presence of high radical concentrations and strong magnetization, the carbon resists graphitization, favoring less delocalized bonds in the form of collapsed CNT, C-onions and nanodiamond. Such instability of radicals and magnetism with graphitic structures has been demonstrated to cause graphitic curvature for fullerenes, CNT and soot formation by the Comprehensive Mechanism of CNT Nucleation and Growth [72]. This impact of magnetic field and its creation of radicals for causing graphitic instability are consistent with the observations of many other researchers. Sun [73] observed that the application of weaker external magnetic field (<1 Tesla) during CNT synthesis eventually caused the CNT process to form amorphous carbon rods rather than CNT. Yokomichi et al. [70] observed that stronger magnetic fields up to 10 T caused greater $C_{70}/C_{60}$ formation during electric arc processes.

Raman [74] realized ring currents in graphite to explain the different magnetic susceptibilities of graphite and diamond. Here it is suggested in this Resolution that in CNT or fullerenes, these ring currents are oriented differently in space relative to planar graphite and this different orientations of ring currents in CNT and fullerenes cause their consequent internal ring---ring diamagnetic repulsions that intrinsically destabilize CNT (fullerenes) in strong magnetic field relative to nanodiamond.  Such ring-ring interactions facilitate the spin transport in CNT [75].  Such effects of the ring currents in CNT are consistent with Kondo Effect [76] and Aharonov-Bohm Effect [757.  Haddon observed temperature and magnetic field dependence of susceptibility of various carbon structures [78].  Roche and Saito [79] observed that magnetism changes electronics of CNT at room temperature.   Unlike graphitic structures, however many researchers have recently reported the stability, polarizing interaction and magnetism of magnetic impurity centers in diamond.  High spin magnetic centers or multi-spin defects have been produced in nanodiamond for the emergence of its ferromagnetism [60,80].  The recent optically observed long lived triplet states [81] in diamond are further evidence of the greater affinity of diamond for spin intermediates and magnetic centers relative to graphitic allotropes.   On the other hand, other researchers have demonstrated the greater incompatibility of radical media with the graphitic lattice.  Enoki [82] demonstrates the strong coupling between such radicals edge states and intrinsic graphitic ring currents.   Just as broken C-C bonds in diamond form lattice C dangling bond, other lattice impurities may have dangling bonds, which the diamond lattice accommodates well relative to the graphitic lattice.  Impurities like Ni, Co, Fe, N and H are accommodated by the diamond in a better way than the graphitic lattice.  Defects are not as tolerated in graphitic structures due to the incompatible interactions of aromatic currents and the neighboring radicals.  The pi cloud is super conducting and the radical is a magnet, so they do not mix in graphite.

## 7. Conclusion:

Diamond is a unique material beyond its extraordinary properties.  The conditions for its growth are rather paradoxical.  The simultaneous need for high volume, high temperature, low pressure conditions to activate its intermediates and for low volume, low temperature, high pressure conditions to organize and stabilize its condensation seem unrealistic over large space and short times.  During the last 50 years limited resolutions of this diamond problem have been provided by the indirect high pressure high temperature methods, the direct super high pressure and high temperature method and the low pressure H vapor deposition.  The magnetic quantum Resolution of this problem given here allows lower volume, lower temperature and atmospheric pressure by correlating radicals, atoms and other intermediates for greater formation, accumulation, organization and correlation of $sp^3$ carbon intermediates over larger space in shorter times for larger, faster single crystal diamond formation.  On the basis of this magnetic quantum fluidic consideration of the reaction trajectories for diamond formation, the Little Effect determines novel electronic dynamics for clarifying the murkiness surrounding carbon integration into a diamond lattice for diamond nucleation and growth.  This magnetic discovery for diamond synthesis has implications for other materials and processes as well.  Use of external magnetization will present a new tool for controlling the reformation of hydrocarbons in conjunction with current high pressure and temperature, catalytic technology.  Using superconducting magnets with such processes such as the Haber process will allow even less costly fixation of atmospheric nitrogen.  The synthesis of boron compounds, singlet oxygen, and halogen oxides such as oxygen difluoride, dioxygen difluoride, and chlorine oxides may be facilitated in external magnetic field.  Many borides form under high pressures and temperatures in the earth's sublayers.   Halogen fluorides are important compounds for future rocket propellants, oxidizing and fluorinating agents.  These halogen oxides cannot be produced directly from halogens and oxygen, so the use of electric arc allows limited formation.  Magnetic methods may resolve the production of these boron compounds and halogen oxides just as the magnetization resolves diamond synthesis.  Strong magnetic effects on some chemical processes present a new era of exploration with great impact on chemistry and physics.

**Acknowledgements**:

In gratitude to GOD.

For my three sons: Reginald Bernard Jr., Ryan Arthur and Christopher Michael.

Special thanks to Prof. Alan Marshall.

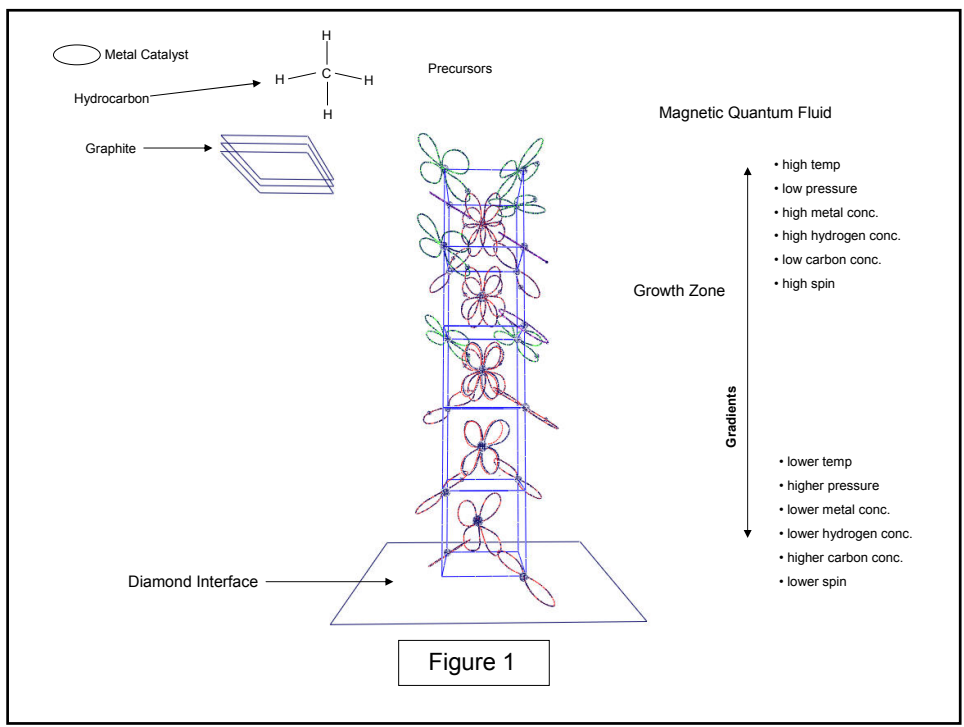

Figure 1

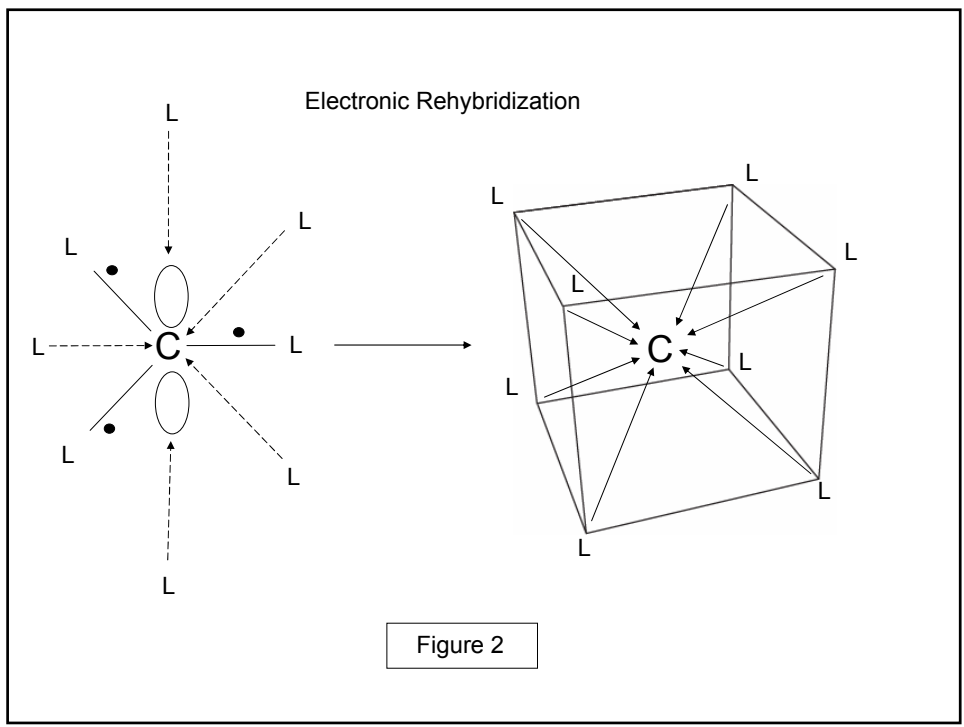

Figure 2

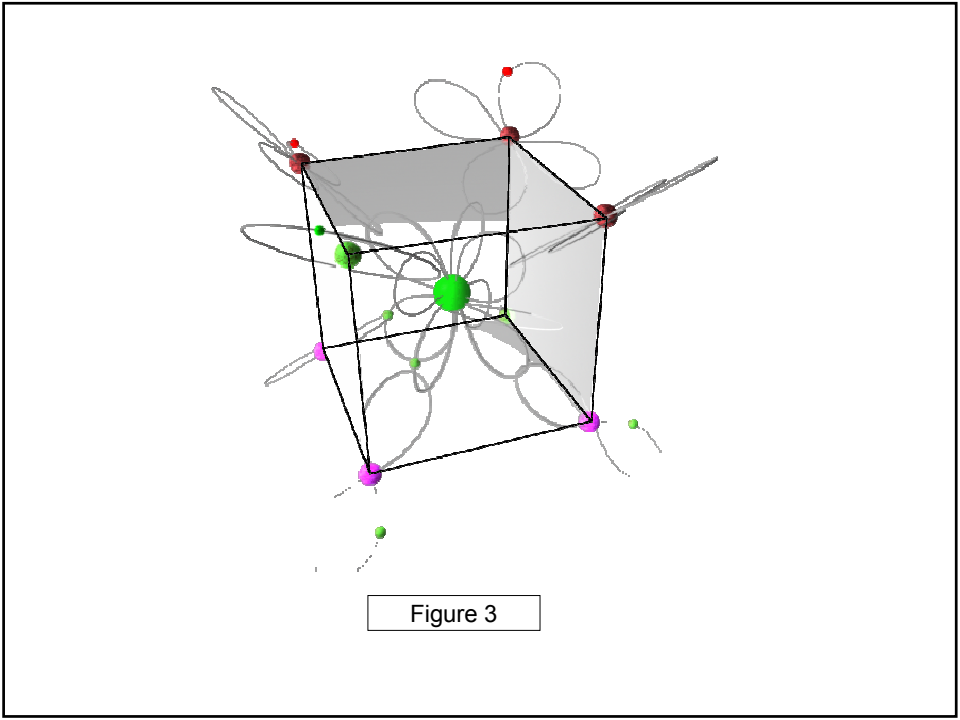

Figure 3

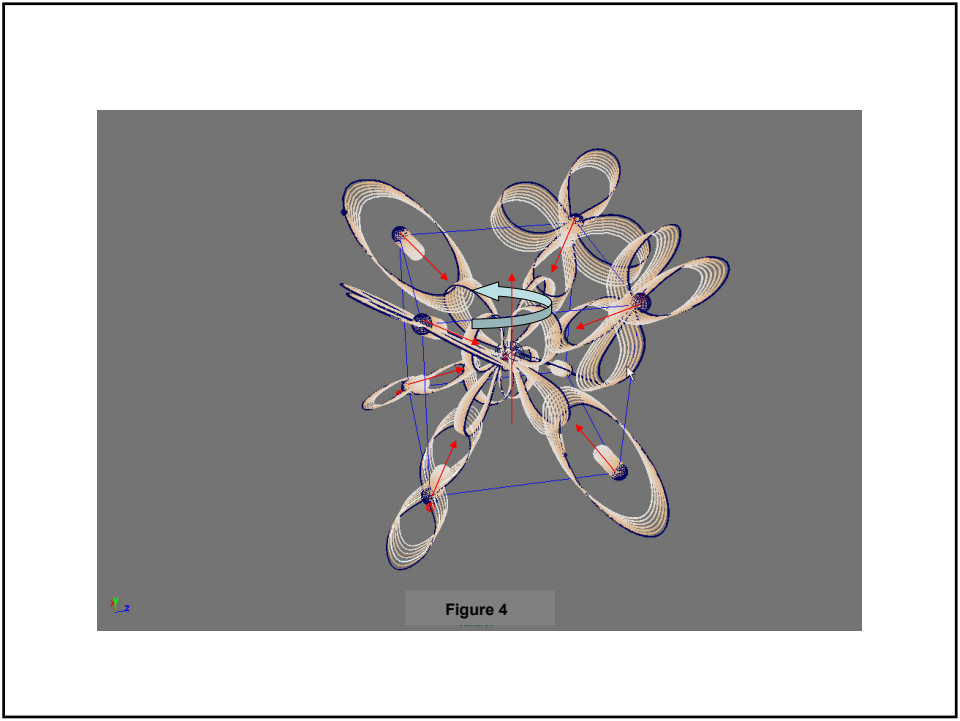

Figure 4